\documentstyle[12pt,psfig]{article}
\newcommand{\half}{\frac 1 2 }

\newcommand{\ie}{{\em i.e.}}
\newcommand{\be}{\begin{eqnarray}}
\newcommand{\ee}{\end{eqnarray}}

\newcommand{\beq}{\begin{equation}}
\newcommand{\eeq}{\end{equation}}

\newcommand{\rmd}{\mbox{d}}
\newcommand{\rme}{\mbox{e}}
\newcommand{\rmi}{\mbox{i}}
\newcommand{\srmi}{\mbox{\scriptsize i}}
\newcommand{\qh}{\mbox{\scriptsize qh}}
\newcommand{\qe}{\mbox{\scriptsize qe}}
\newcommand{\oqh}{\mbox{\scriptsize 1qh}}
\newcommand{\oqe}{\mbox{\scriptsize 1qe}}
\newcommand{\tqh}{\mbox{\scriptsize 2qh}}
\newcommand{\tqe}{\mbox{\scriptsize 2qe}}

\def\p{\partial}

\catcode`\@=11
\newcommand{\ccite}[1] {\@ifundefined{b@#1}{\bf ?}{\@nameuse{b@#1}}}
\catcode`\@=12

\begin{document}
\vspace*{1cm}
\centerline{\Large\bf Numerical study of charge and statistics}
\vspace*{0.25\baselineskip}
\centerline{\Large\bf of Laughlin quasi-particles} 
\vspace*{-45 mm}
\vskip 55mm
\centerline{\bf Heidi Kj{\o}nsberg$^{\dagger}$$^{\ddagger}$
and Jan Myrheim$^{\star}$  } 
\vskip 3mm
\centerline{$^{\dagger}$Department of Physics, University of Oslo,
P.O.\ Box 1048 Blindern}
\centerline{N--0316 Oslo, Norway}
\centerline{e-mail: heidi.kjonsberg@fys.uio.no}
\centerline{$^{\star}$Department of Physics,
The Norwegian University of Science and Technology, NTNU}
\centerline{N--7034 Trondheim, Norway}
\centerline{e-mail: myrheim@phys.ntnu.no}
\vskip 15mm
\def\fk{\mbox{ $f_K$} }
\centerline{\bf ABSTRACT}
\vspace{.5cm}
We present numerical calculations of the charge and statistics,
as extracted from Berry phases, of the Laughlin quasi-particles,
near filling fraction $1/3$, and for system sizes of up to
200 electrons. For the quasi-holes our results confirm that
the charge and statistics parameter are $e/3$ and $1/3$, respectively.
For the quasi-electron charge we find a slow convergence towards
the expected value of $-e/3$, with a finite size correction for
$N$ electrons of approximately $-0.13e/N$.
The statistics parameter for the quasi-electrons has no well defined
value even for 200 electrons, but might possibly
converge to $1/3$. Most noteworthy, it takes on the same
sign as for the quasi-holes, due to terms that have previously
been ignored.
The anyon model works well for the quasi-holes, but requires
singular two-anyon wave functions for modelling two Laughlin
quasi-electrons.
\medskip
\vskip 3mm

\vfil
\noindent
$^{\ddagger}$Supported by The Norwegian Research Council.

\eject

\section{Introduction}  

It is now widely accepted that the fractional quantum Hall effect
arises due to quasi-particle excitations created when
the filling fraction of the lowest Landau level moves away from its
preferred values. The quasi-particles bind
to impurities and thereby ensure that the plateaus in the
conductivity are formed. Laughlin examined the filling fractions $1/m$,
with $m$ an odd integer, and argued that the quasi-particles have
charge $\pm e/m$ ($-e$ is the electron charge) \cite{laughlin}.
He showed that these values for the charge imply a conductivity
plateau at the value $e^2/(mh)$, as observed.
He also offered explicit trial wave functions describing the
ground state as well as the quasi-particle excitations for
the $1/m$ case. Later, 
Haldane \cite{haldane} and Halperin \cite{halperin} examined the
hierarchy structure of the Hall states, and suggested that the
quasi-particles obey fractional statistics, \ie\ that they are anyons
\cite{jonm_jan}.  

Arovas, Schrieffer and Wilczek derived the charge and statistics of the
Laughlin quasi-holes in Ref.~\cite{arovas}. They
examined  the Berry phase \cite{berry} corresponding to one
quasi-hole encircling the origin and interpreted this as an
Aharonov-Bohm phase \cite{berry,AharonBohm}. The charge was then found to be
$e/m$, in confirmation of Laughlin's result. They also considered a pair
of quasi-holes encircling one another, and related the two-particle
contribution of the Berry phase to the quasi-hole charge. Interpreting
this two-particle contribution as an
anyon interchange phase they found the anyon statistics parameter to
have the value $1/m$, equal to the fraction of the elementary charge.
The Laughlin quasi-electrons were examined along the same lines in
Ref.~\cite{arovas2}. Within the approximations used, the results imply
that the charge and statistics parameter have the values $-e/m$
and $-1/m$, respectively.

The statistics satisfied by the quasi-particles in the quantum Hall
system was also examined in Ref.~\cite{john_can},
where the exclusion statistics parameter \cite{haldane2} was
considered. The results did not rely on any specific trial wave
function but rather on state counting based on numerical
simulations for interacting electrons on a sphere.
The value of the one-dimensional exclusion statistics
parameter was found to be $1/m$ in the case of quasi-holes,
while the value $2-1/m$ was found for the quasi-electrons. Both
cases were examined near to the magic filling fraction $1/m$
with $m=3$.

The exclusion statistics parameter is in principle the same parameter
as one reads from the Berry phase, although with an opposite sign
for the quasi-electrons, because their charge is negative.
The relation is briefly discussed in the Appendix below.
Thus the predictions for the anyon parameter, based on
the numerical results for the exclusion statistics parameter,
would be $1/m$ for quasi-holes and $-2+1/m$ for quasi-electrons.
The values $1/m$ and $-2+1/m$ for the anyon parameter define
of course the same particle statistics, but we distinguish between
them here in the way we define the correspondence between
quasi-particles and anyons. Thus, in the case of quasi-electrons,
$1/m$ and $-2+1/m$ would represent the same species of anyons,
but different anyon states, the $1/m$ state singular and
the $-2+1/m$ state non-singular.

It is interesting to note that the numerical results for
the exclusion statistics parameter of realistic quasi-holes and
quasi-electrons are easily interpreted in terms of anyons of
positive and negative charge, respectively, with non-singular
wave functions, and with the same statistics in the two cases.
That quasi-holes and quasi-electrons should have the same statistics,
is also what one would expect if one regards them as antiparticles
of each other.

The anyon representation for the quasi-particles, first suggested by
Laughlin \cite{laughplasma,laughfrac}, was considered in more detail in
Ref.~\cite{jonm_hei}. A careful examination of the quasi-electron
case revealed that the Berry connection calculated from the Laughlin
wave function actually contains
some terms not earlier considered in the literature. These
extra terms arise because the inverse of the operator used to create a
quasi-electron is not simply the complex conjugate of the inverse
quasi-hole operator. There is no obvious argument for neglecting them,
and this motivates a closer examination of whether these
wave functions really represent excitations carrying the charge and
statistics parameter characterizing the physical quasi-electrons
in the quantum Hall system. There are no similar problems associated
with the Laughlin quasi-hole wave functions.

The purpose of the work presented here was to compute the Berry phases
for the Laughlin quasi-particles in order to derive their charge
and statistics, following the ideas of Arovas et al.\ \cite{arovas}.
The calculations were performed both for quasi-holes and
quasi-electrons, with special attention to the latter,
at the filling fraction $1/3$.
For a small number of electrons ($\leq 8$) we could integrate
exactly, but for larger and more realistic systems
we had to use Monte Carlo integration with importance sampling
according to the Metropolis algorithm \cite{laughplasma}.

In the case of a single quasi-hole as well as for a pair of quasi-holes,
we find that the Berry phases computed for systems with up to 200
electrons behave as expected. Thus the charge $e/3$ and the statistics
parameter $1/3$ for the quasi-holes are both confirmed.

For a single quasi-electron the extra terms affect the Berry phase
so as to imply a charge that for a system with up to 200 electrons
is slightly below $-e/3$, and also the boundary
effects are larger than for the quasi-hole. For a pair of
quasi-electrons, the computations give a definite short distance
behaviour of the statistics phase, but at distances larger than
about one magnetic length, which is roughly the size of a quasi-electron,
the phase does not settle down to an asymptotic value, as one would
like it to. Moreover, to the extent that the statistics parameter is
a meaningful quantity, it has the same sign as for the quasi-holes,
in clear contradiction to earlier results for the Laughlin
quasi-particles.  
This sign would correspond to singular two-anyon wave functions,
and hence a negative exclusion statistics parameter.

Within the last year there has been reported
direct experimental evidence for the fractional charge of the
quasi-particles in the fractional quantum Hall system
\cite{experiment}. According to a conjecture proposed in
Ref.~\cite{kivelson}, fractional charge implies fractional
statistics. Thus our investigation of the charge and statistics of
the Laughlin quasi-particles has some current interest
for comparison with the real quantum Hall system.

\section{Charge and Statistics From Berry Phases}

Consider a system of $N$ particles (electrons) of charge
$-e$ ($e>0$), moving in the $(x,y)$ plane, with a magnetic field
${\bf B}= B {\bf e}_{z}$ ($B>0$), perpendicular to the plane.
The position of a particle may be described by
the complex coordinate $z=(x+iy)/(\sqrt{2}l)$, where
$l=1/\sqrt{eB}$ is the magnetic length. We set $\hbar,c=1$.

Due to the magnetic field there is an Aharonov-Bohm phase
associated with the propagation of a charged
particle around a closed loop \cite{AharonBohm}.
If the path is a circle of radius $\rho=\sqrt{2}l r$,
right handed relative to the direction of the external
magnetic field, and if the particle has charge $q$,
this phase is given by
\beq
\gamma =\pi \rho^2 Bq = 2\pi\,\frac{q}{e}\, r^2\;.
\eeq

Consider a quasi-particle (quasi-electron or quasi-hole) excitation,
of charge $q$, localized at the position $z_0$ and described
by the normalized wave function
\beq
\Psi_{z_0}(z_1, _{\cdots}\! ,z_N)=
\langle z_1, _{\cdots}\! ,z_N\,|\,z_0\rangle\;.
\eeq
If this quasi-particle is moved around a closed loop,
there arises the so called Berry phase, which is
the integral along the path of the Berry connection \cite{berry}.
Let the path be a circle around
the origin, parametrized as $z_0= r\rme^{\srmi\phi}$ with $\phi$
running from $0$ to $2\pi$. The Berry connection is then defined by
\beq
\frac{\rmd\beta_1}{\rmd\phi} =
\rmi\,\langle z_0|\,\p_\phi\,|z_0 \rangle\;. \label{bcon1}
\eeq

The charge $q$ of the quasi-particle is now determined by setting
the Berry phase equal to the Aharonov-Bohm phase corresponding to the
same path \cite{arovas}. In this way $q$ is expressed in terms of
parameters of the fundamental particles (electrons) in the system.
In a finite system the charge defined in this way will depend on
the distance $r$ from the origin, but it is reasonably
well defined as long as the $r$ dependence is small for circles
that are well inside the system.

Suppose the quasi-particle is described by a normalized $N$-particle
state of the form 
\beq
|z_0\rangle = {1\over\sqrt{I_1}}\sum_{l=0}^{\infty}
{z_0}^l\,a_l\,|l\rangle\;.
\label{tilst}
\eeq
Here $a_l$ are expansion coefficients, $|l\rangle$
are orthonormal basis states, and $I_1$ is introduced for
normalization,
\beq
I_1=\sum_{l=0}^{\infty} r^{2l} |a_l|^2\;.
\eeq
Then the Berry connection depends on $r$ but not on $\phi$,
so that the Berry phase is
$\beta_1(2\pi)=2\pi\,\rmd\beta_1/\rmd\phi$.
The expression for the charge becomes especially simple,
and is given by the $r$ dependence of $I_1$ as
\be
\frac{q}{e}=\frac{1}{ r^2}\,\frac{\rmd\beta_1}{\rmd\phi}
           =-\frac{\rmd}{\rmd r^2}\ln I_1\;.
\ee

Now suppose there are two quasi-particle excitations simultaneously,
located symmetrically about the origin at the positions $\pm z_0$.
The parametrization $z_0= r\rme^{\srmi\phi}$ then describes
a counterclockwise interchange of the
quasi-particles if we let $\phi$ run from $0$ to $\pi$.
If we assume that the two quasi-particle state is described by
a state analogous to (\ref{tilst}),
\beq
|z_0,-z_0\rangle = {1\over\sqrt{I_2}}\sum_{l=0}^{\infty}
{z_0}^l\,b_l\,|l'\rangle\;,  \label{tilst2}
\eeq
then a Berry connection $\rmd\beta_2/\rmd\phi$ corresponding to
this interchange may be defined in the same way as (\ref{bcon1}),
\beq
\frac{\rmd\beta_2}{\rmd\phi} =
\rmi\,\langle z_0,-z_0|\,\p_\phi\,|z_0,-z_0\rangle\;. \label{bcon2}
\eeq
We subtract the single-particle contributions due to
the magnetic field, and define
\beq
-\nu=
\frac{1}{\pi}\,(\beta_2(\pi) - 2\beta_1(\pi))=
\frac{\rmd}{\rmd\phi}\,(\beta_2 - 2\beta_1)=
-r^2\frac{\rmd}{\rmd r^2}\,(\ln I_2 - 2\ln I_1)\;.
\label{statpar}
\eeq
We will refer to $\nu$ as the ``anyon parameter'', since
the residual Berry phase $-\nu\pi$ can be identified with
minus the statistics phase of the particles.

As a technical point, note that Eq.~(\ref{statpar}) gives $\nu$
as the small difference of two quantities that are large for large
$r$. When computing $\nu$ by the Monte Carlo method, it is important
to evaluate both integrals $I_1$ and $I_2$ simultaneously, using
the same sample of random points, since
the statistical errors will then be correlated and tend to cancel.

The Laughlin quasi-particles of the quantum Hall system are
described by states of the same form as in Eq.~(\ref{tilst}),
or by states of this form with $z_0$ replaced by $z_0^*$.
In the next section we give the explicit expressions
for the wave functions. For now it suffices to know that the
expansion states corresponding to $|l\rangle$ in (\ref{tilst})
are orthogonal, and the same is true for $|l'\rangle$ in
(\ref{tilst2}), which are different from $|l\rangle$.
A non-normalized quasi-electron state is
expanded as a polynomial in $z_0$, whereas a quasi-hole state
is expanded in $z_0^*$ instead of $z_0$.
The complex conjugation shows up as a difference in sign
in the relation between the Berry connection and
the normalization factors $I_1$ and $I_2$.
We use the notation
$I^{\oqh},I^{\tqh}I^{\oqe},I^{\tqe}$ for the normalization
factors for one and two quasi-holes and one and two quasi-electrons,
respectively, with $N$ electrons in the system. The
following expressions for the charges and
anyon parameters of these quasi-particles then result,
\be
\frac{q^{\qh}}{e} &=&   \frac{\rmd}{\rmd r^2}\ln I^{\oqh}\;,
\label{chqh} \\  
\frac{q^{\qe}}{e} &=&  -\frac{\rmd}{\rmd r^2}\ln I^{\oqe}\;,
\label{chqe} \\   
\nu^{\qh}         &=&  -r^2\frac{\rmd}{\rmd r^2}\,
(\ln I^{\tqh}-2\ln I^{\oqh})\;,
\label{stqh}  \\  
\nu^{\qe}         &=&   r^2\frac{\rmd}{\rmd r^2}\,
(\ln I^{\tqe}-2\ln I^{\oqe})\;.
\label{stqe}
\ee

This shows immediately that the charge is positive for
the quasi-holes and negative for the quasi-electrons.
The sign of the anyon parameter in the two cases is
not obvious directly from these formulae.
As discussed in the Appendix, the anyon model predicts
that $\nu^{\qh}\geq 0$ and $\nu^{\qe}\leq 0$,
when only non-singular two-anyon wave functions are considered,
whereas the ranges $-1<\nu^{\qh}<0$ and $0<\nu^{\qe}<1$
correspond to singular two-anyon wave functions.

\section{Computational methods}

\subsection*{One quasi-hole}

The non-normalized $N$-electron wave function for one quasi-hole at
the position $z_0$ is
\be
\Psi_{z_0}^{\oqh}(z,z^*)
=\psi_0\,\Delta^{*m}\prod_{i=1}^N(z_i^*-z_0^*)
=\psi_0\,\Delta^{*m}\sum_{k=0}^N (-z_0^*)^k S_{k,N}(z^*)\;. 
\ee
We simplify our notation by writing $z=(z_1,_{\cdots}\!,z_N)$.
Here $1/m$ is the filling fraction, $\psi_0$ is
the harmonic oscillator ground state wave function,
\be
\psi_0=\rme^{-\half\sum_{i=1}^N |z_i|^2}\;,
\ee
and $\Delta$ is the Vandermonde determinant,
\be
\Delta=\prod_{j<k}(z_j -z_k)\;.
\ee
The elementary symmetric polynomials $S_{k,N}$ \cite{lang} are
implicitly defined by the above expansion, which brings
the quasi-hole state explicitly to the form shown in Eq.~(\ref{tilst}),
except for the substitution $z_0\to z_0^*$.
For computations we use the following recursion relation, 
\be
S_{k,N}(z)=z_N\,S_{k,N-1}(z)+S_{k-1,N-1}(z)\;,
\label{sam}
\ee
with $S_{0,0}=1$ and all other $S_{k,0}=0$.

The normalization integral, which determines
the charge according to Eq.~(\ref{chqh}), is given by
\be
I^{\oqh}( r^2)
&=& \int\rmd^{2N}\!z\,|\Psi_{z_0}^{\oqh}|^2\\
&=& \int\rmd^{2N}\!z\,{\psi_0}^2\,|\Delta|^{2m}
\prod_{k=1}^N |z_k-z_0|^2   \label{qhplasma}\\
&=& \sum_{k=0}^N  r^{2k} \int\rmd^{2N}\!z 
\,{\psi_0}^2\,|\Delta|^{2m}\,|S_{k,N}(z)|^2 \label{qhexp}\\
&=& \sum_{k=0}^N  r^{2k} I_k^{\oqh}\;. \label{rekke}
\ee

Since we want the logarithmic derivative with respect to
$ r^2$, we need compute only the ratios $I_k^{\oqh}/I_N^{\oqh}$.
We are able to do this analytically only for very small $N$.
For larger $N$ we do Monte Carlo
integration by the Metropolis algorithm \cite{laughplasma}.
This method works well in particular because
\be
I_N^{\oqh}=\int\rmd^{2N}\!z\,{\psi_0}^2\,|\Delta|^{2m}\;,
\ee
and the integration measure
$\rmd^{2N}\!z\,{\psi_0}^2\,|\Delta|^{2m}$ is common to all
the integrals $I_k^{\oqh}$. We interpret
${\psi_0}^2\,|\Delta|^{2m}$ as a Boltzmann factor $\rme^{-\beta V}$
of classical statistical mechanics, thinking of $\beta$ as
the ``inverse temperature'' and $V$ as the ``potential energy'', with
\be
\beta V=\sum_i|z_i|^2-2m\sum_{i<j}\ln|z_i-z_j|\;.
\ee
The classical system is a plasma with two-dimensional Coulomb
repulsion, in an external harmonic oscillator potential
\cite{laughplasma}. The factor $\prod_{k=1}^N |z_k-z_0|^2$,
which we do not include in the integration measure,
represents in this plasma analogy a Coulomb repulsion of the electrons
from the hole position $z_0$.

We compute the ratio $I_k^{\oqh}/I_N^{\oqh}$ as the time average
of $|S_{k,N}(z)|^2$ over a time sequence of $N$-electron
configurations generated by the following random dynamics.
We loop through all electrons, and for the $i$-th electron we generate
a random ``trial jump'' $z_i\to z_i+\Delta z_i$ in such a way that
$\Delta z_i$ and $-\Delta z_i$ are equally probable.
The trial step is accepted or rejected depending on whether
the Boltzmann factor $\rme^{-\beta\,\Delta V}$, where $\Delta V$
is the change in potential energy, is larger or smaller than
a random number generated uniformly between 0 and 1.
This procedure gives the desired distribution of configurations,
because the dynamics satisfies the principle of detailed balance:
the ratio of probabilities for jumping from $A$ to $B$ or from
$B$ to $A$ is the Boltzmann factor $\rme^{-\beta(V(B)-V(A))}$.

\subsection*{One quasi-electron}

Now consider a quasi-electron at the position $z_0$.
Laughlin's proposed wave function is a polynomial in $z_0$,
\be
\Psi_{z_0}^{\oqe}(z,z^*)
=\psi_0\left(\prod_{i=1}^N(\p_{z_i^*}-z_0)\right)\Delta^{*m}
=\psi_0\sum_{k=0}^N (-z_0)^k S_{k,N}(\p^*)\,\Delta^{*m}\;,
\ee
where $\p^*=(\p_{z_1^*},_{\cdots}\!,\p_{z_N^*})$.
The normalization integral can be written in several different ways.
We may expand it as a polynomial in $r^2$,
\be
I^{\oqe}( r^2)
&=& \int\rmd^{2N}\!z\,|\Psi_{z_0}^{\oqe}|^2  \\
&=& \sum_{k=0}^N  r^{2k} \int\rmd^{2N}\!z
\,{\psi_0}^2\,|S_{k,N}(\p)\,\Delta^m|^2  \label{exp}\\
&=& \sum_{k=0}^N  r^{2k} I_k^{\oqe}\;, \label{pol}
\ee
or rewrite it by partial integration as
\be
I^{\oqe}( r^2)=\int\rmd^{2N}\!z\,{\psi_0}^2\,|\Delta|^{2m}
\prod_{k=1}^N (|z_k-z_0|^2-1)\;.
\label{plasma}
\ee
The last form shows clearly the difference between the quasi-hole and
quasi-electron cases.
If we could neglect the $-1$ in each factor $|z_k-z_0|^2-1$,
then the integrals would be identical, and the charges would be
the same, just with opposite signs. This approximation seems hard
to justify, since the average number of electrons within unit
distance from an arbitrary point $z_0$, assuming constant density
within the electron droplet, is $\pi/m$, which is close to one
if $m=3$. A more likely possibility would be that the normalization
integral does indeed change, but only by a factor which is
mainly independent of $ r$, and which therefore does not change
the Berry phase.

Note that the integrand in Eq.~(\ref{plasma}) may even be negative,
in spite of the fact that the integrand of the original normalization
integral is explicitly non-negative. This can happen because
the derivation of Eq.~(\ref{plasma}) involves $2N$ partial
integrations. The cancellation between positive and negative
contributions will cause some loss of precision when this form
of the integral is used for numerical evaluation.

Before proceeding we want to comment on the origin of the extra $-1$.
Remember that the wave function of one particle in a magnetic field,
in the lowest Landau level, has the form $f(z^*)\,\rme^{-\half|z|^2}$,
where $f$ is an analytic function. The projection onto the lowest
Landau level of the operator $z$ is the operator $\p_{z^*}$ acting on
the space of functions analytic in $z^*$ \cite{girvin}.
This is seen by partial integration, which gives that
\beq
\int\rmd^2\!z\,\rme^{-zz^*}(g(z^*))^*\p_{z^*} f(z^*)
=\int\rmd^2\!z\,\rme^{-zz^*}(g(z^*))^* z f(z^*) \label{proj} 
\eeq
for two analytic functions $f$ and $g$.
Partial integration also gives that
\be
\label{type}
\int\rmd^2\!z\,\rme^{-zz^*}
 ((\p_{z^*}-z_0) g(z^*))^* (\p_{z^*}-z_0) f(z^*)
=\int\rmd^2\!z\,\rme^{-zz^*}(g(z^*))^*\,(|z-z_0|^2-1) f(z^*)\;.
\ee  
In the projection onto the lowest Landau level there is an operator 
ordering problem, because of the fundamental fact that 
$[z^*,\p_{z^*}]=-1$.
Thus, the factor $|z-z_0|^2$ in the quasi-hole normalization integral
corresponds to the ordering $(z-z_0)(z^*-z_0^*)$, whereas
the factor $|z-z_0|^2-1$ in the quasi-electron integral
corresponds to $(z^*-z_0^*)(z-z_0)$.

We may use Eq.~(\ref{plasma}) to compute the integrals $I_k^{\oqe}$
defined in Eq.~(\ref{exp}), by expanding
\beq
\prod_{n=1}^N (|z_n-z_0|^2-1)
=\sum_{k,l=0}^N a_{k,l}^N\,{z_0}^k\,{z_0^*}^l\;.
\eeq
The coefficients $a_{k,l}^N$ satisfy the following recursion
relation,
\beq
a_{k,l}^n=(|z_n|^2-1)\,a_{k,l}^{n-1}-z_n^*\,a_{k-1,l}^{n-1}
-z_n\,a_{k,l-1}^{n-1}+a_{k-1,l-1}^{n-1}\;,
\eeq
with $a_{0,0}^0=1$ and all other $a_{k,l}^0=0$.
Only the diagonal coefficients $a_{k,k}^N$ give nonzero contributions
to the integral, but in order to compute them we have to compute some,
though not all, of the off-diagonal coefficients $a_{k,l}^n$.
Note that $a_{k,k}^n$ is real, and more generally,
$a_{l,k}^n=(a_{k,l}^n)^*$.

\subsection*{Two quasi-holes}

For two
quasi-holes at the positions $z_a$ and $z_b$ the wave function is 
\beq
\Psi_{z_a,z_b}^{\tqh} = \psi_0\,\Delta^{*m}\,
\prod_{i=1}^N(z_i^*-z_a^*)(z_i^*-z_b^*)\;.
\eeq
The normalization integral also for this state takes the
form of a probability distribution for a system with two-dimensional
Coulomb interactions, 
\beq
I_{z_a,z_b}^{\tqh} =  \int\rmd^{2N}\!z\,{\psi_0}^2\,|\Delta|^{2m}
\prod_{k=1}^N |(z_k - z_a)(z_k - z_b)|^2\;,  \label{kvusym}
\eeq
which is analogous to Eq.~(\ref{qhplasma}). If we
let the quasi-holes be located symmetrically around the origin,
$z_a=-z_b=z_0$, then the state is on the form of Eq.~(\ref{tilst2}),
with $z_0\to z_0^*$,
and the normalization integral can be expressed as
\beq
I^{\tqh}( r^2)=\sum_{k=0}^N  r^{4k} I_k^{\tqh}\;.
\eeq
Here
\beq
I_k^{\tqh} = \int\rmd^{2N}\!z\,{\psi_0}^2\,|\Delta|^{2m}
\,|S_{k,N}(z_1^2,_{\cdots}\!,z_N^2)|^2 
\eeq
are again normalization integrals for certain angular momentum
eigenstates. The elementary symmetric polynomials now depend on $z^2$
instead of $z$. Numerical calculation of these integrals is just as
straightforward as in the case of a single
quasi-hole, since the recursion formula in Eq.~(\ref{sam}) is valid
with the substitution $z\rightarrow z^2$.

\subsection*{Two quasi-electrons}

Two quasi-electrons at $z_a$ and $z_b$ are described by the wave
function
\beq
\Psi_{z_a,z_b}^{\tqe} = \psi_0
\left(\prod_{i=1}^N(\p_{z_i^*}-z_a)
(\p_{z_i^*}-z_b)\right)\Delta^{*m}\;,
\eeq
which yields, by partial integration, the normalization integral
\be
I_{z_a,z_b}^{\tqe}=
\int\rmd^{2N}\!z\,{\psi_0}^2\,|\Delta|^{2m}
\prod_{k=1}^N \left(|z_k-z_a|^2|z_k-z_b|^2
-\left|2z_k-z_a-z_b\right|^2+2\right).
\label{kvusymII}
\ee
Comparing with the quasi-hole integral in Eq.~(\ref{kvusym}),
we see that for two quasi-electrons there are correction terms, 
like in the case of a single quasi-electron.

We choose to calculate Berry phases for two quasi-electrons located
symmetrically around the origin, with $z_a=-z_b=z_0$, which
again ensures that our state is of the form given in Eq.~(\ref{tilst2}).
The normalization integral needed to find the statistics parameter
according to Eq.~(\ref{stqe}), takes the form
\beq
I^{\tqe}( r^2) = \sum_{k=0}^N  r^{4k} I_k^{\tqe}\;,
\eeq
where
\beq
I_k^{\tqe}=\int\rmd^{2N}\!z\,{\psi_0}^2\,
|S_{k,N}(\p_{z_1}^2,_{\cdots}\!,\p_{z_N}^2)\Delta^m|^2\;.
\eeq
However, it is again simpler to compute $I_k^{\tqe}$ by first
integrating partially to eliminate the derivatives
and then expanding
\beq
\prod_{n=1}^N (|{z_n}^2-{z_0}^2|^2-4|z_n|^2+2)
=\sum_{k,l=0}^N b_{k,l}^N\,{z_0}^{2k}\,{z_0^*}^{2l}\;.
\eeq
The coefficients $b_{k,l}^N$ satisfy the recursion relation
\beq
b_{k,l}^n=(|z_n|^4-4|z_n|^2+2)\,b_{k,l}^{n-1}
-z_n^{*2}\,b_{k-1,l}^{n-1}
-{z_n}^2\,b_{k,l-1}^{n-1}+b_{k-1,l-1}^{n-1}\;,
\eeq
with $b_{0,0}^0=1$ and all other $b_{k,l}^0=0$.
Again it is only the diagonal coefficients $b_{k,k}^N$ that give
nonzero contributions to the integral.

\section{Results}

In this section we will present results from the Monte Carlo
calculations described above.
We have plotted, as functions of the dimensionless radius
$r$ of a circular loop, the charge and statistics parameter
for Laughlin quasi-holes and quasi-electrons, extracted from
Berry phases according to Eqs.~(\ref{chqh}), (\ref{chqe}),
(\ref{stqh}) and (\ref{stqe}).

The case we have studied is $m=3$, i.e.\ $1/3$ filling of
the lowest Landau level. For the quasi-holes we find Berry
phases that imply the charge $e/3$ and the statistics parameter
$1/3$, in confirmation of the results from Ref.~\cite{arovas}.

For the quasi-electrons our results are not nearly as unambiguous.
The difference as compared to the quasi-hole case is due to
the extra terms in the normalization integrals, as discussed above,
and for the first time in Ref.~\cite{jonm_hei}.
These terms affect the value of the quasi-electron charge, making
it slightly larger than $e/3$ in absolute value.
Nevertheless the charge is close to the expected
value of $-e/3$, and it seems plausible that the deviation is
a finite size effect, vanishing in the limit of
infinitely many electrons.

The statistics parameter of the quasi-electrons is where
the difference shows up most clearly. In particular, the statistics
parameter is found to have the same sign as for quasi-holes,
and not the opposite sign as would be the case if the extra
terms were unimportant. Furthermore, as seen in the figures
presented below, there is no clear indication in our numerical
results that the statistics parameter will converge either
to $1/3$ or to any other value as the number of electrons increases.

\begin{figure}[htb]
\begin{center}
\hspace*{-9mm}
\mbox{\psfig{figure=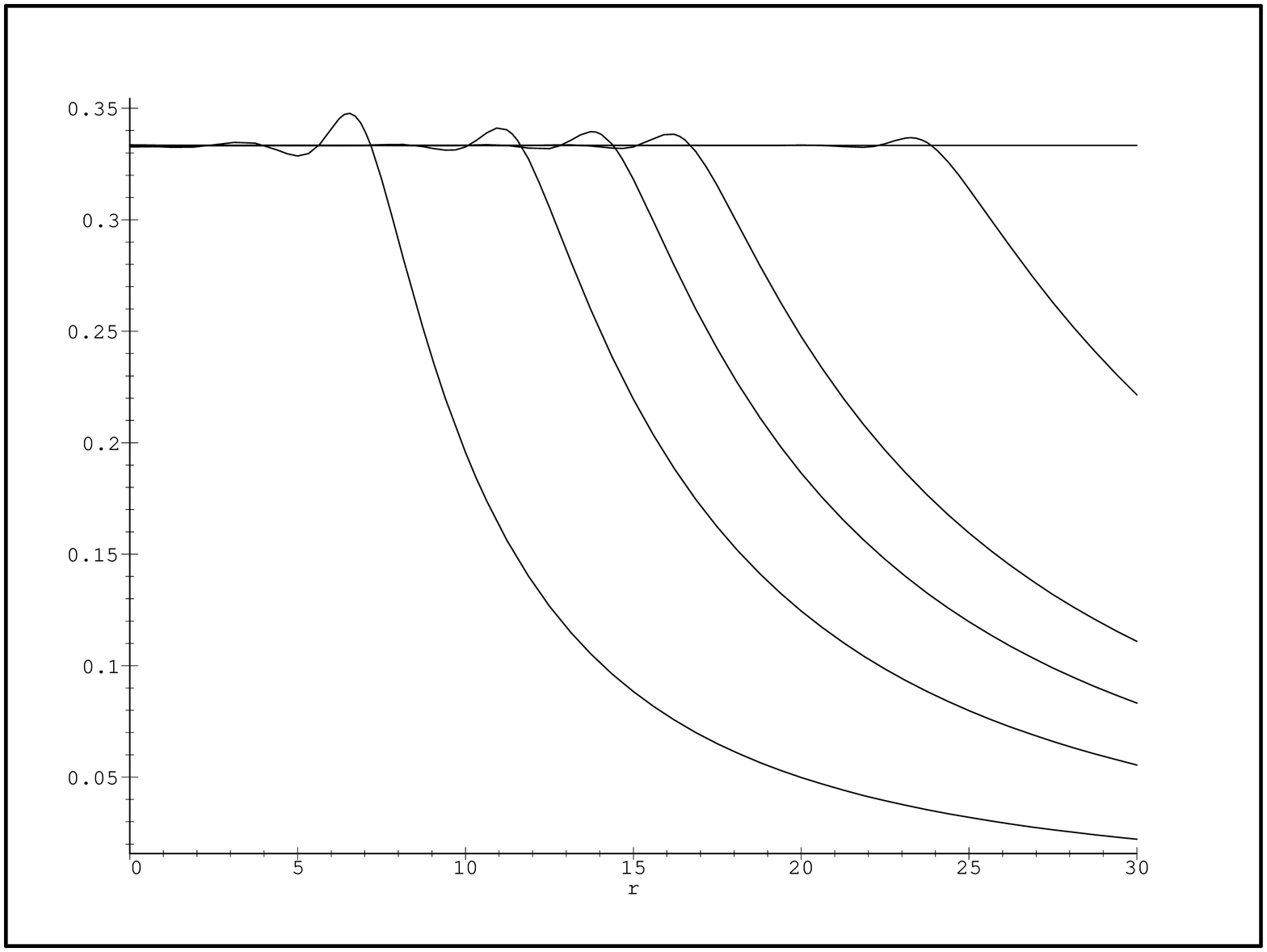,angle=270,height=14cm}}
\end{center}
\caption[]{
The quasi-hole charge $q^{\qh}/e$,
Eq.~(\ref{chqh}), as a function of $r$,
the quasi-hole distance from the origin. The curves are, from left
to right, for 20, 50, 75, 100 and 200 electrons.
The horizontal line is $1/3$.}
\label{fig1}
\end{figure}

Consider first the bulk value of the quasi-hole
charge $q^{\qh}$ as given by Eq.~(\ref{chqh}).
In Fig.~\ref{fig1} the quantity $q^{\qh}/e$ is shown as a function of $r$,
the dimensionless distance from 
the origin to the quasi-hole.
For filling fraction $1/m$ the radius of a circular electron
droplet with $N$ electrons is approximately $\sqrt{mN}$.
With $m=3$ and for the values of $N$ considered here
this means that the droplet boundaries are at $r\approx 7.7$,
$12.2$, $15.0$, $17.3$ and $24.5$. Fig.~\ref{fig1} then clearly
illustrates that as
long as the quasi-hole is well inside the droplet, its charge is well
defined and equal to $1/3$ of the absolute value of the
electron charge. This result is in accordance with the bulk value
proposed in Ref.~\cite{laughlin}, and later calculated in
Ref.~\cite{arovas}. We also note that
the boundary effects weaken as the number of electrons grows. 

\begin{figure}[htb]
\begin{center}
\hspace*{-9mm}
\mbox{\psfig{figure=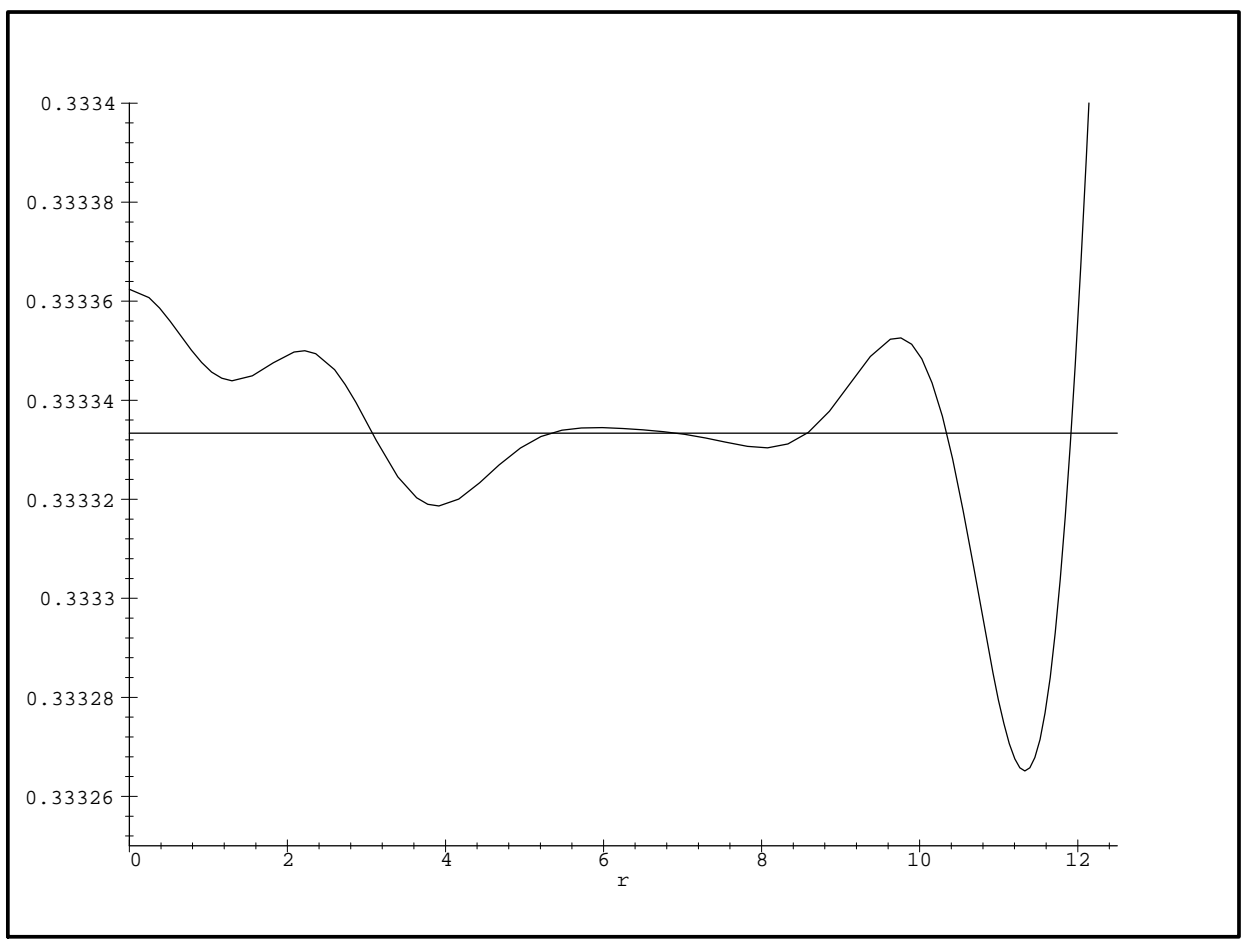,angle=270,height=14cm}}
\end{center}
\caption[]{
The quasi-hole charge, $q^{\qh}/e$, compared to $1/3$,
for 100 electrons.}
\label{fig2}
\end{figure}

In Fig.~\ref{fig2} we show a more detailed picture of
the 100 electron case, where we have collected
a statistics of 409 million points in the Monte Carlo
integration. The bulk value, which we define as the value for
not too small and not too large radius, is seen to take
the expected value of $1/3$ to at least
four decimal digits. We consider the deviations from $1/3$ to be
not statistically significant, this we checked in some cases
by splitting the data in two or more parts and then plotting
curves for each partial data set. The deviations from $1/3$ at
large $r$ are clearly edge effects due to the finite system size.

The deviations at small $r$ we believe to be also finite size
effects that can be understood. In fact, the logarithmic derivative
of $I^{\oqh}(r^2)$ is determined at $r=0$ by only two coefficients
$I^{\oqh}_0$ and $I^{\oqh}_1$, whereas for medium values of $r$
it is a kind of average over many coefficients.
Each coefficient may be subject to fluctuations, due to
the finite Monte Carlo statistics and perhaps also due to the finite
system size. Such fluctuations are more likely to be averaged out
in quantities depending on many coefficients.

\begin{figure}[htb]
\begin{center}
\hspace*{-9mm}
\mbox{\psfig{figure=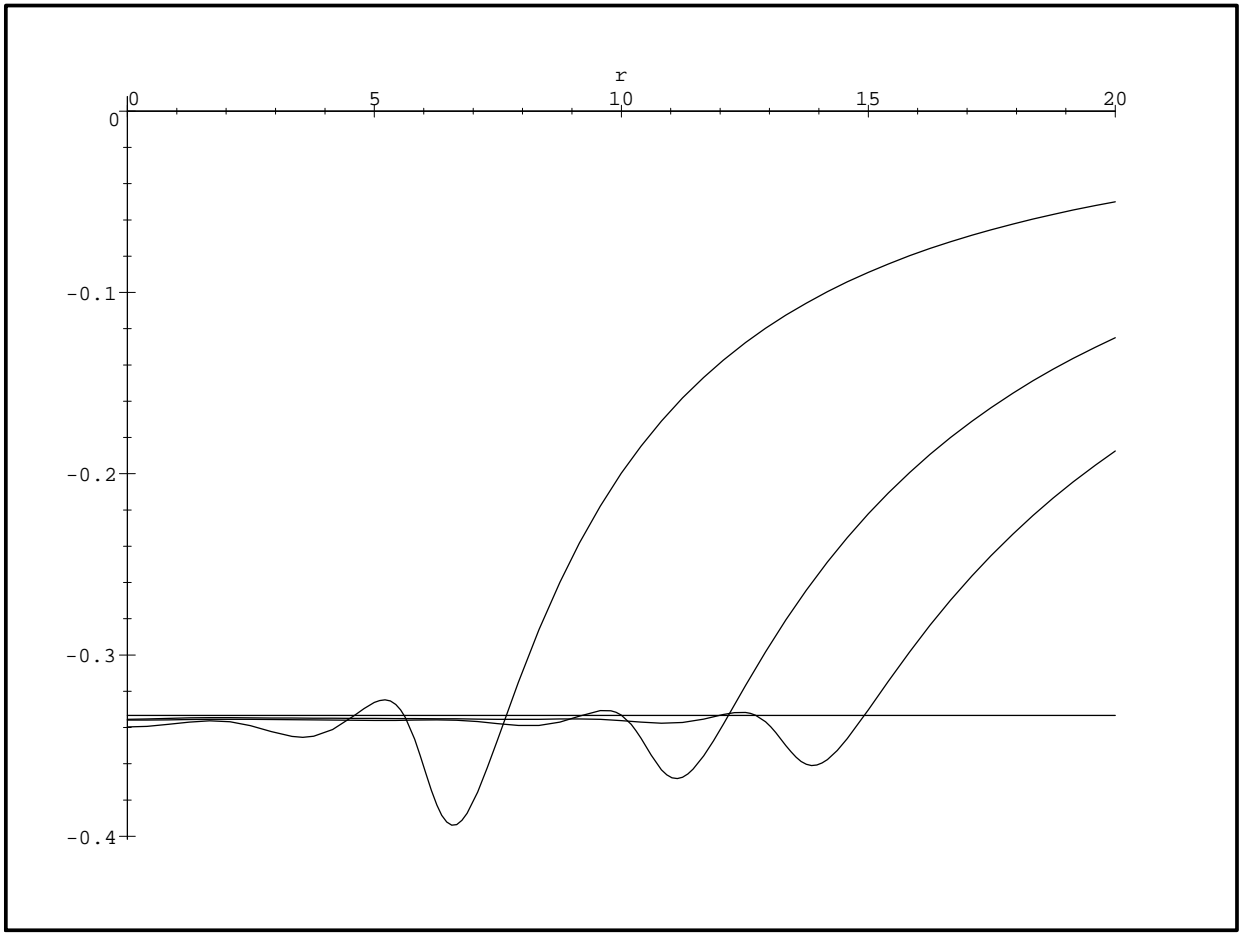,angle=270,height=14cm}}
\end{center}
\caption[]{
The quasi-electron charge $q^{\qe}/e$,
Eq.~(\ref{chqe}), as a function of $r$,
the quasi-electron distance from the origin. The figure presents
results for, from left to right, 20, 50 and 75 electrons.
The horizontal line is $-1/3$.}
\label{fig3}
\end{figure}

Fig.~\ref{fig3} shows the charge of the quasi-electron in terms of
the absolute value of the electron charge, $q^{\qe}/e$, which is given
by Eq.~(\ref{chqe}) and is expected to have the value $-1/3$.
As for the quasi-holes, the charge is well defined in the bulk, and
the boundary effects decrease with increasing electron number. However,
the bulk value is slightly lower than $-1/3$.
This result fits together with the common belief
expressed elsewhere in the literature only if the deviation vanishes
with increasing electron number, which may well happen, although
our present data are not conclusive evidence.
Comparing Figs.~\ref{fig1} and \ref{fig3} we see
that the edge effect due to the finite size of the electron
droplet is larger in the case of a quasi-electron than it is for the
quasi-hole. Both of these features, the deviation from $-1/3$ and
the larger edge effects, are due to the extra terms in the
normalization integrals.

\begin{figure}[htb]
\begin{center}
\hspace*{-9mm}
\mbox{\psfig{figure=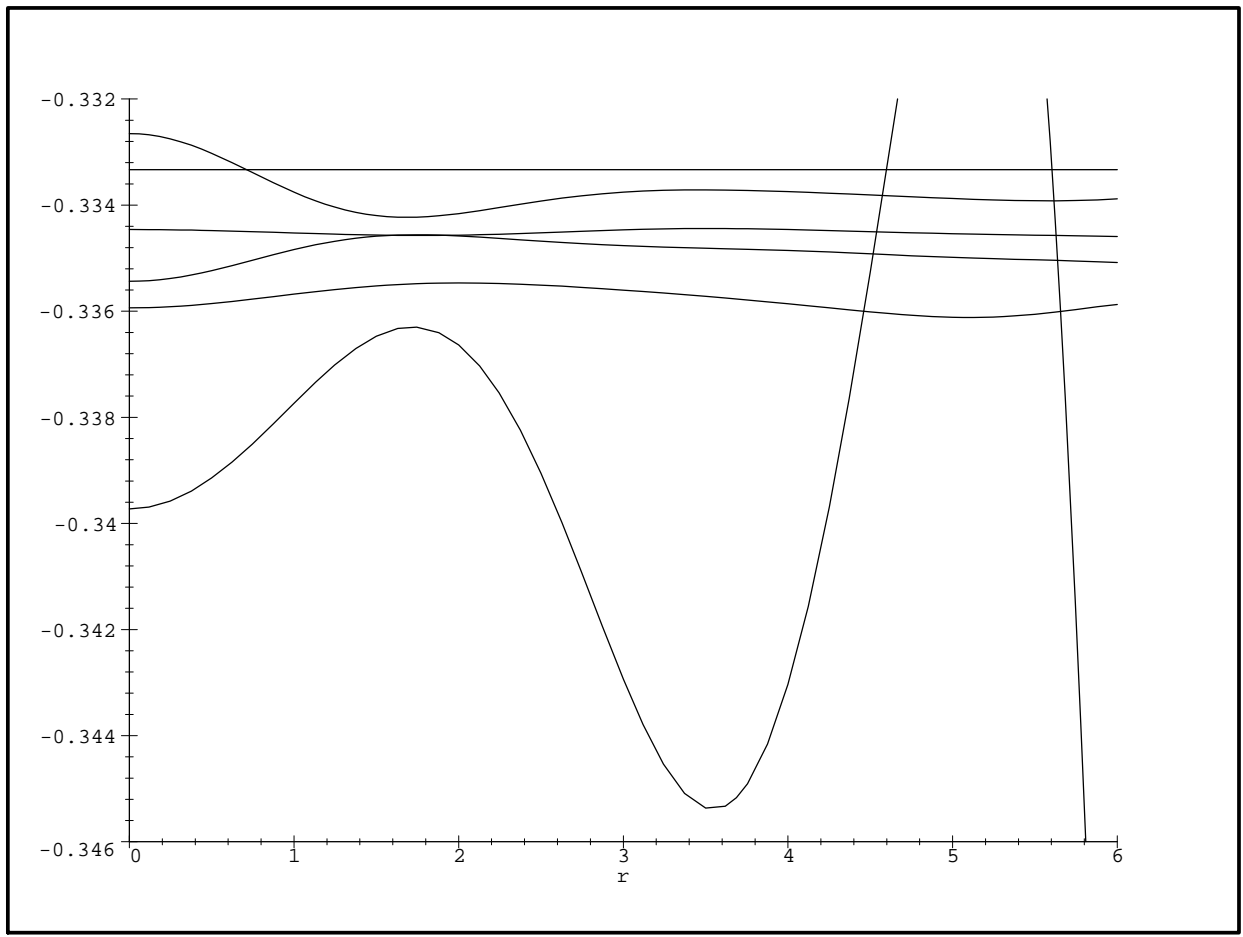,angle=270,height=14cm}}
\end{center}
\caption[]{
Quasi-electron charge, $q^{\qe}/e$.
The number of electrons, from the lowest lying curve and up, is
20, 50, 75, 100 and 200. The highest curve is the constant $-1/3$.}
\label{fig4}
\end{figure}

Fig.~\ref{fig4} shows $q^{\qe}/e$ for small distances,
$r\leq 6$, including also curves computed for 100 and 200 electrons.
The results shown may be parametrized roughly as
\beq
\frac{q^{\qe}}{e} = -\frac{1}{3}-\frac{0.13}{N}\;.
\eeq
To the extent that this parametrization is valid, the conclusion
is that the deviation from $-1/3$ is a finite size effect.

\begin{figure}[htb]
\begin{center}
\hspace*{-9mm}
\mbox{\psfig{figure=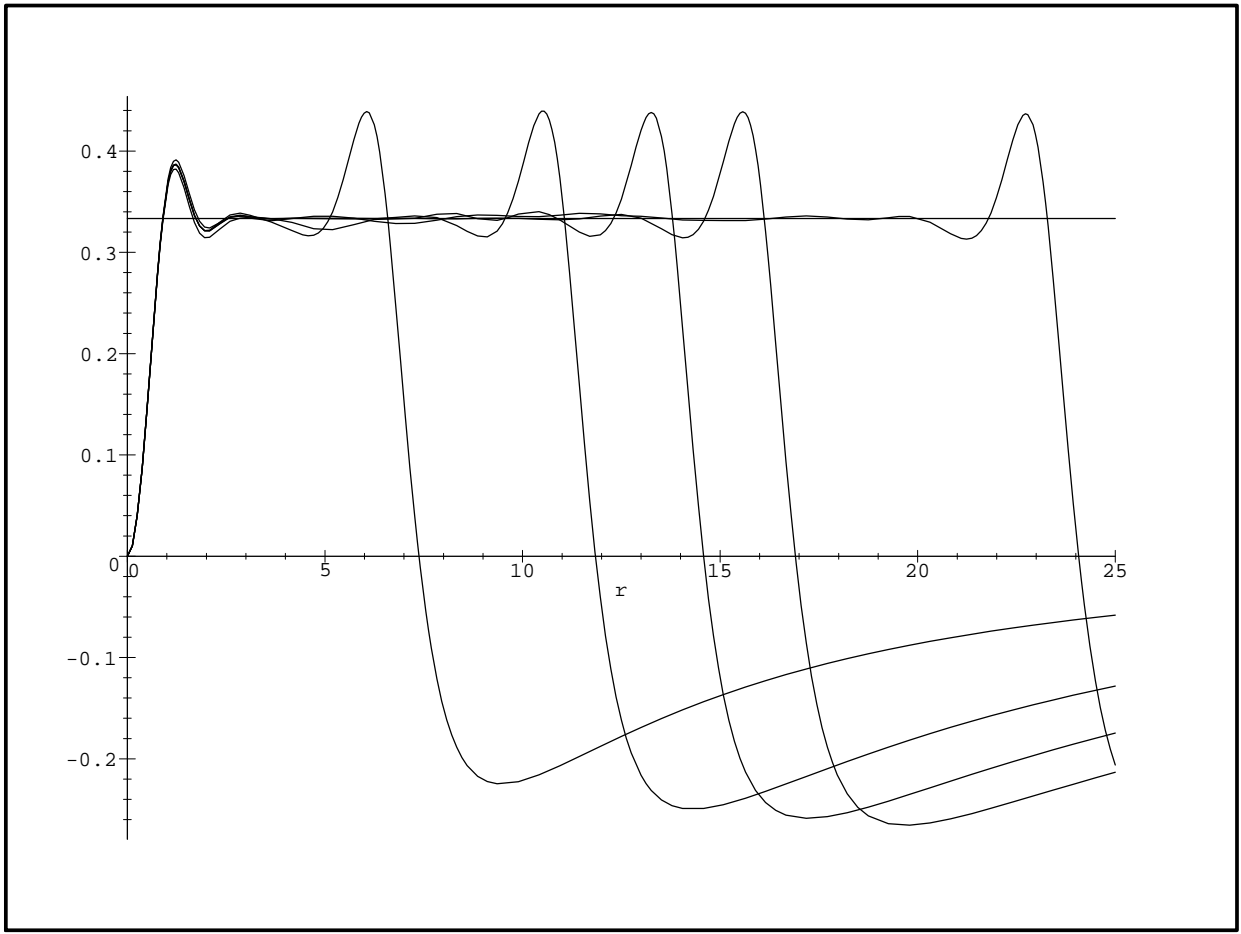,angle=270,height=14cm}}
\end{center}
\caption[]{
The quasi-hole statistics parameter $\nu^{\qh}$, Eq.~(\ref{stqh}), as a
function of $r$, half the distance between the two quasi-holes. The
curves are, from left to right, for 20, 50, 75, 100 and 200
electrons, and the horizontal line is $1/3$.}
\label{fig5}
\end{figure}

We focus next on the statistics parameter $\nu^{\qh}$ for the Laughlin
quasi-holes, Eq.~(\ref{stqh}). Fig.~\ref{fig5} shows this quantity as
a function of $r$, half the distance between the two holes.
When the two quasi-holes are well inside the electron 
droplet and also reasonably well separated, the quantity $\nu^{\qh}$
is seen to have a rather well defined constant value of $1/3$.
These two characteristics together make it meaningful to state that
the quasi-holes have the anyon statistics parameter $1/3$.
Another general feature is that when the two quasi-holes
are close to the edge of the electron droplet the curves show a dip,
followed by a peak which does not change size or shape when the
number of electrons changes, in contrast to the behaviour seen in
Fig.~\ref{fig1}.

Fig.~\ref{fig5} also demonstrates how the curves go to zero when
the two quasi-holes overlap, \ie\ $r\rightarrow 0$. In
Fig.~\ref{fig6} this small relative distance behaviour is shown
on a larger scale for five different electron numbers,
and compared to the Berry phases calculated from two
different states in the system of two ideal anyons.
The three lowest lying curves are, from bottom and up, for
75 electrons, then coinciding curves
for 20, 50 and 100 electrons, and finally for 200 electrons.
We conclude that the behaviour at small relative
distance is due to the finite size of the
quasi-holes themselves and is not affected by the number of electrons
in the system.

The two highest curves in Fig.~\ref{fig6} do not refer to
the quantum Hall system, but offer 
an explicit comparison of features of the Laughlin quasi-holes to the 
case of ideal, point like, anyons, as discussed in the Appendix.
In Ref.~\cite{jonm_hei} it was pointed out that the first order
correction to the constant, 
asymptotic value of the two quasi-hole Berry phase 
favours the projected anyon position eigenstate rather than the 
coherent state of the su(1,1) algbra as the localized anyon 
state that best fits to the Laughlin quasi-hole state. This conclusion 
was based on the plasma analogy, saying essentially that
the corrections to perfect screening
should fall off exponentially rather than algebraically.
It is confirmed by Fig.~\ref{fig6}, where the curve representing
the coherent state lies highest, and
the curve representing the projected anyon state indeed lies closer 
to the quasi-hole curves.

The fact that the curves in Fig.~\ref{fig6} go to zero for $r\to 0$,
can be understood in the anyon model as due to the impossibility
of constructing states with the point like anyons completely
localized, using only states belonging to the lowest Landau level.
The parameter $z$, or $z^*$, defining the anyon states
we consider, is different from the relative position of the anyons,
and moving $z$ to $-z$ around a circle of radius $r$ is not
the same as interchanging the anyon positions. Hence, in the limit
$r\to 0$, no anyon statistics phase is picked up.

In Fig.~\ref{fig7} we show the bulk behaviour of $\nu^{\qh}$, 
as calculated for 100 electrons. Again we take this as
a confirmation, to at least three decimal digits, of
the known result of $1/3$ for the quasi-hole statistics parameter.

Our final topic is what we have up to now referred to as the statistics
parameter of the quasi-electrons, \ie\ $\nu^{\qe}$ in Eq.~(\ref{stqe}).
Fig.~\ref{fig8} shows this quantity for the cases of
20, 50, 75, 100 and 200 electrons. The 100 and 200 electron curves
had to be cut at $r=8$ and $r=6$, respectively, because of
numerical problems. This figure faces us with the question of
whether the Laughlin quasi-electrons can be said to have a well defined
statistics parameter at all. In our
opinion, the term ``statistics parameter'' is justified only if the
quantity denoted is independent of the distance
between the two quasi-particles as long as they neither overlap
nor are close to the edge of the electron droplet. We
have seen that this criterion is well satisfied by the quantity
$\nu^{\qh}$, Eq.~(\ref{stqh}), but Fig.~\ref{fig8} reveals that
it definitely does not hold for $\nu^{\qe}$, Eq.~(\ref{stqe}),
for the system sizes we have considered.

We would like to
stress that if we had neglected the extra terms in the integrals
in Eqs.~(\ref{plasma}) and (\ref{kvusymII}) as compared to
Eqs.~(\ref{qhplasma}) and (\ref{kvusym})
then this strange behaviour would not have appeared, and the statistics
parameter for the quasi-electrons would have been equal
to minus the statistics parameter of the quasi-holes.
Thus, the fact that the computed $\nu^{\qe}$ and $\nu^{\qh}$ have the same
sign, is alone proof that the extra terms are essential.

Fig.~\ref{fig9} shows the same curves on an enlarged scale.
It also includes two curves that offer an explicit
comparison to the case of ideal, point like anyons, for small $r$
these lie below and to the right of the curves
referring to the quantum Hall system. Note that the five
quasi-electron curves depend only slightly on the number 
of electrons for small $r$, and thus offer a signature 
of the Laughlin quasi-electrons themselves. Note also that
all the curves start out negative at small $r$ before becoming positive.

The two curves referring to the anyon system are calculated from states
that are singular but normalizable. In fact, since the calculated right hand
side of Eq.~(\ref{stqe})
is positive, the only possible way to extract
an anyon parameter is to make a correspondence with such
singular states in the ideal anyon system. Fig.~\ref{fig9} shows that
the coherent state is closest to the quasi-electron curves for small
$r$, whereas the projected position eigenstate gives a better fit
for large $r$.

Fig.~\ref{fig9} indicates that the bulk value of $\nu^{\qe}$ may
possibly become well defined (\ie\ constant) in the thermodynamic 
limit. But if so, it is not obvious that the limiting value
would be $1/3$.

\section{Concluding remarks}

To summarize, our numerical study has
nicely confirmed the results earlier presented in the literature,
that the quasi-holes at filling fraction $1/3$ have charge $e/3$,
and have anyon parameter and exclusion statistics parameter both
equal to $1/3$. 

If it were true that the normalization factors $I_1$ and $I_2$ were
the same for the Laughlin quasi-electrons and quasi-holes, then
the Berry phases for the quasi-electrons would be the same as for the
quasi-holes, only with opposite signs. This would then lead us to the
numerical results $q^{\qe}=-e/3$ and $\nu^{\qe}=-1/3$, whereas the
exclusion parameter would be $\nu_1^{\qe} = +1/3$.
 
However, for the quasi-electrons we have seen that for systems with 
up to 200 electrons the charge is affected by the extra terms 
discussed above, so as to give a bulk value that is slightly lower 
than the expected $-e/3$. The deviation is quite small and may well
vanish for infinitely many electrons, but we do find it interesting
to note that the quasi-electrons, which 
have a size of the order one magnetic length \cite{laughplasma},
seem to be affected by the edge even when it is as much as
ten magnetic lengths away.

For two quasi-electrons encircling
one another our results differ dramatically from those earlier
proposed in the literature. 
The extra terms affect the Berry phase such that it is
unclear whether one can extract any statistics parameter at all,
even in the thermodynamic limit. What we can conclude is that
the the statistics parameter, if at all meaningful, has a sign
which is opposite to what it would have if the extra terms 
were unimportant. It may possibly converge to $1/3$ in the
thermodynamic limit, even though we find no strong indication that
it does. If the Laughlin quasi-electrons have an anyon parameter
$\nu^{\qe}=1/3$, it means that the anyon wave functions are singular,
and the exclusion statistics parameter is
$\nu_1^{\qe} = - \nu^{\qe} = -1/3$. This result fits in
with the naive argument saying that if the quasi-holes and quasi-electrons
are anti-particles, they should have the same anyon parameter. 
However, it is in not in agreement with the value $2-1/3$ found
from state counting, which corresponds to non-singular anyon wave
functions.

\section*{Acknowledgements}

We would like to thank J.M.\ Leinaas, K.\ Olaussen, G.S.\ Canright and
T.H.\ Hansson for helpful discussions and useful suggestions.

\appendix
\section{Berry phases of the two-anyon states}
\label{appen}

In order to model the Laughlin quasi-particles as anyons
of statistics parameter $\nu$ in the lowest
Landau level, it is necessary to establish a correspondence with
specific states in the anyon system. There are two different
two-anyon states that are natural to consider as
states that maximally localize each of the particles
\cite{jonm_hei,jonm_hans_jan}. They are the coherent state
of the su(1,1) algebra and the anyon coordinate
eigenstate projected onto the lowest Landau level.
These states are identical in the fermion and boson cases,
but not in general.

Assume that the magnetic field is $B$ and the anyon charge is $e/m$,
with $e>0$ and $B>0$.
The relative angular momentum of the two anyons, in the lowest Landau
level, have eigenvalues $2k+\nu$ with $k$ integer.
Let $|k,\nu\rangle$ be the corresponding orthonormal eigenstates,
and let $|u\rangle$ be the position eigenstates, with $u$ the complex
relative coordinate. Asymptotically as $|u|\to 0$ we have that
\beq
\langle u\mid k,\nu\rangle\propto u^{2k+\nu}\;.
\eeq
Usually one requires that $2k+\nu\geq 0$, so that the angular
momentum eigenstates are non-singular. But if $-1<2k+\nu<0$
the corresponding state would be normalizable, although singular,
and one may include it.

The coherent state can be written
\beq
\mid \! z,\nu\rangle = {\cal N}_z \sum_{k=0}^\infty
\,\frac{z^{*2k}}{m^k\,\sqrt{k!\Gamma(k+\nu+\half)}}\,
\mid \! k,\nu\rangle\;,
\label{koa}
\eeq
whereas the projected position eigenstate is
\beq
\mid \! z,\nu\rangle = {\cal N}'_{z} \sum_{k=0}^\infty
\,\frac{2^kz^{*2k}}{m^k\,\sqrt{\pi \Gamma(2k+\nu+1)}}\,
\mid \! k,\nu\rangle\;.
\label{anya}
\eeq
Here ${\cal N}_z$ and ${\cal N}'_{z}$ are
normalization factors, and $z$ is a parameter labelling
the states, such that $2z$ is to be interpreted as
the relative position of the two anyons, measured in units
of $\sqrt{2}$ times the magnetic length $1/\sqrt{eB}$.
This scaling of distances corresponds to the convention
used for the quantum Hall system in this paper.
The only restriction on the statistics parameter $\nu$
in these formulae is that $\nu\geq 0$, if
the states are required to be non-singular, or that
$\nu>-1$, if singular but normalizable states are accepted.
But note that the sums leave out one or more of the non-singular
angular momentum eigenstates when $\nu\geq 2$.

For the statistics parameter as measured by the Berry phase
we find the expression
\beq
C_{\nu}(r) = \frac{2}{m}\,r^2 - r^2 \frac{d}{dr^2}
\ln\!\left(\frac{I_{\nu-\half}(\frac{2}{m}r^2)}{r^{2\nu -1}}\right)
\label{kober} 
\eeq
for the coherent state, with $I_{\nu - 1/2}$ a modified
Bessel function, and
\beq
A_{\nu}(r) = \frac{2}{m}\,r^2 - r^2 \frac{d}{dr^2}
\ln\!\left(\sum_{l=0}^\infty
\frac{(\frac{2}{m}r^2)^{2l}}{\Gamma(2l+\nu+1)}\right) \label{anber}
\eeq
for the projected anyon position eigenstate.
In both cases, the one-particle contribution $-2r^2/m$ has been
subtracted, and the asymptotic value as $r\to\infty$ is $\nu$.

\begin{figure}[htb]
\begin{center}
\hspace*{-9mm}
\mbox{\psfig{figure=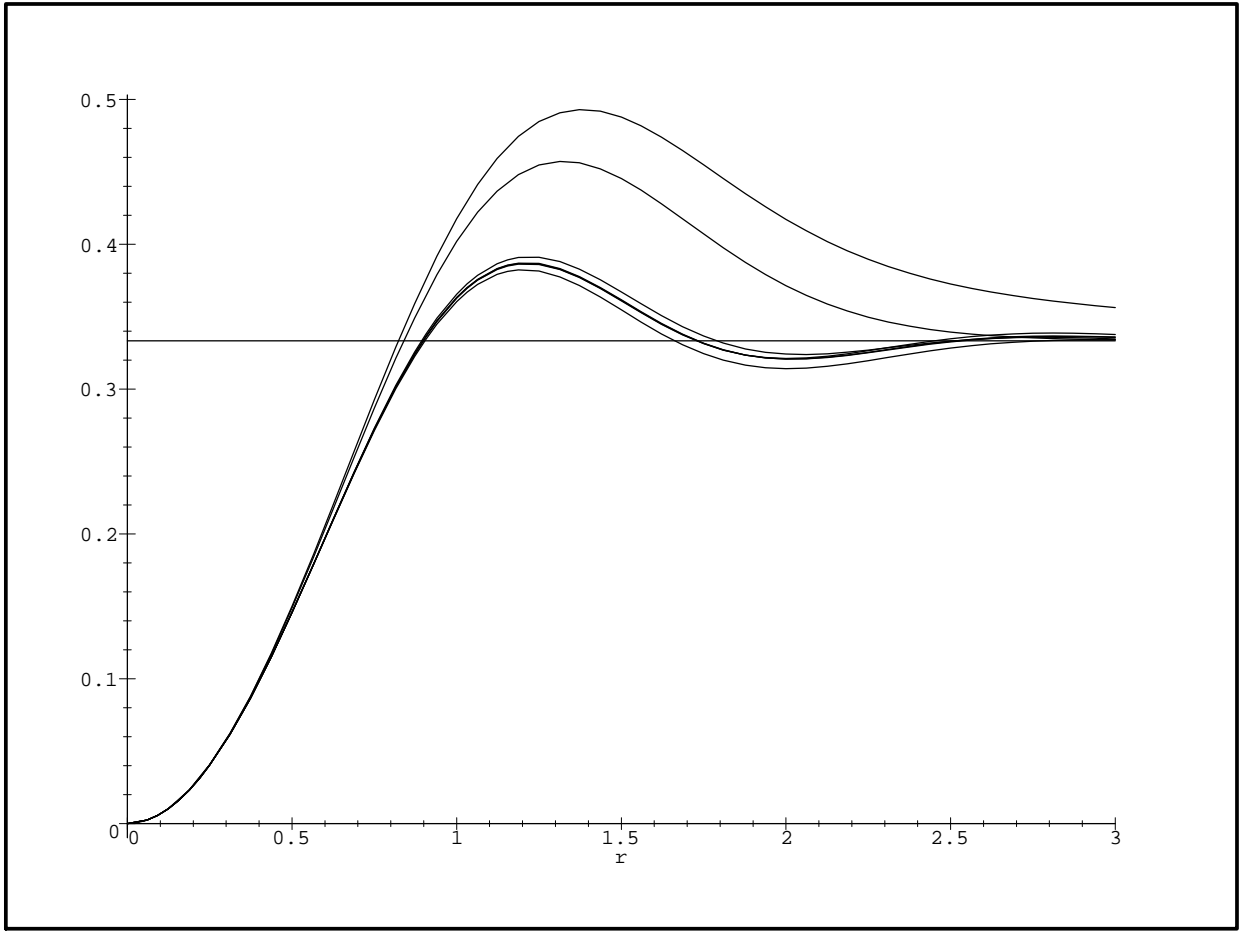,angle=270,height=14cm}}
\end{center}
\caption[]{
Small relative distance behaviour of $\nu^{\qh}$, and Berry phases
calculated from two different localized states in the system of two
anyons. The lowest lying curve is for 75 electrons, then follows
a common curve for the three cases of 20, 50 and 100 electrons, and
the third curve is for 200 electrons. Somewhat higher lies the
Berry phase curve calculated from the anyon position eigenstate
projected onto the lowest Landau level, and even higher the one
calculated from the coherent state of the su(1,1) algebra.
See the Appendix for details.}
\label{fig6}
\end{figure}

\begin{figure}[htb]
\begin{center}
\hspace*{-9mm}
\mbox{\psfig{figure=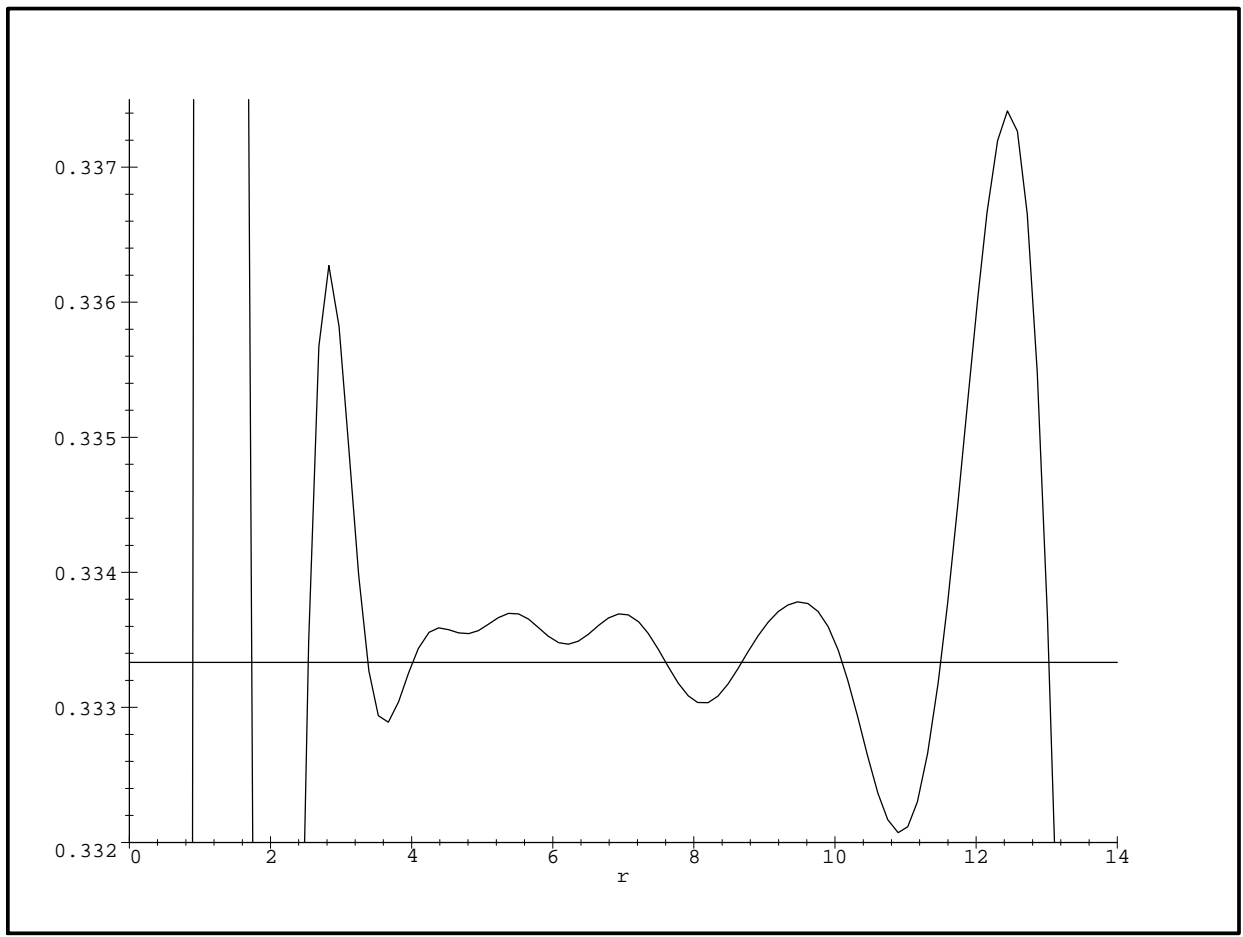,angle=270,height=14cm}}
\end{center}
\caption[]{
Quasi-hole statistics parameter $\nu^{\qh}$ for 100 electrons,
compared to $1/3$, emphasizing the bulk behaviour.}
\label{fig7}
\end{figure}

\begin{figure}[htb]
\begin{center}
\hspace*{-9mm}
\mbox{\psfig{figure=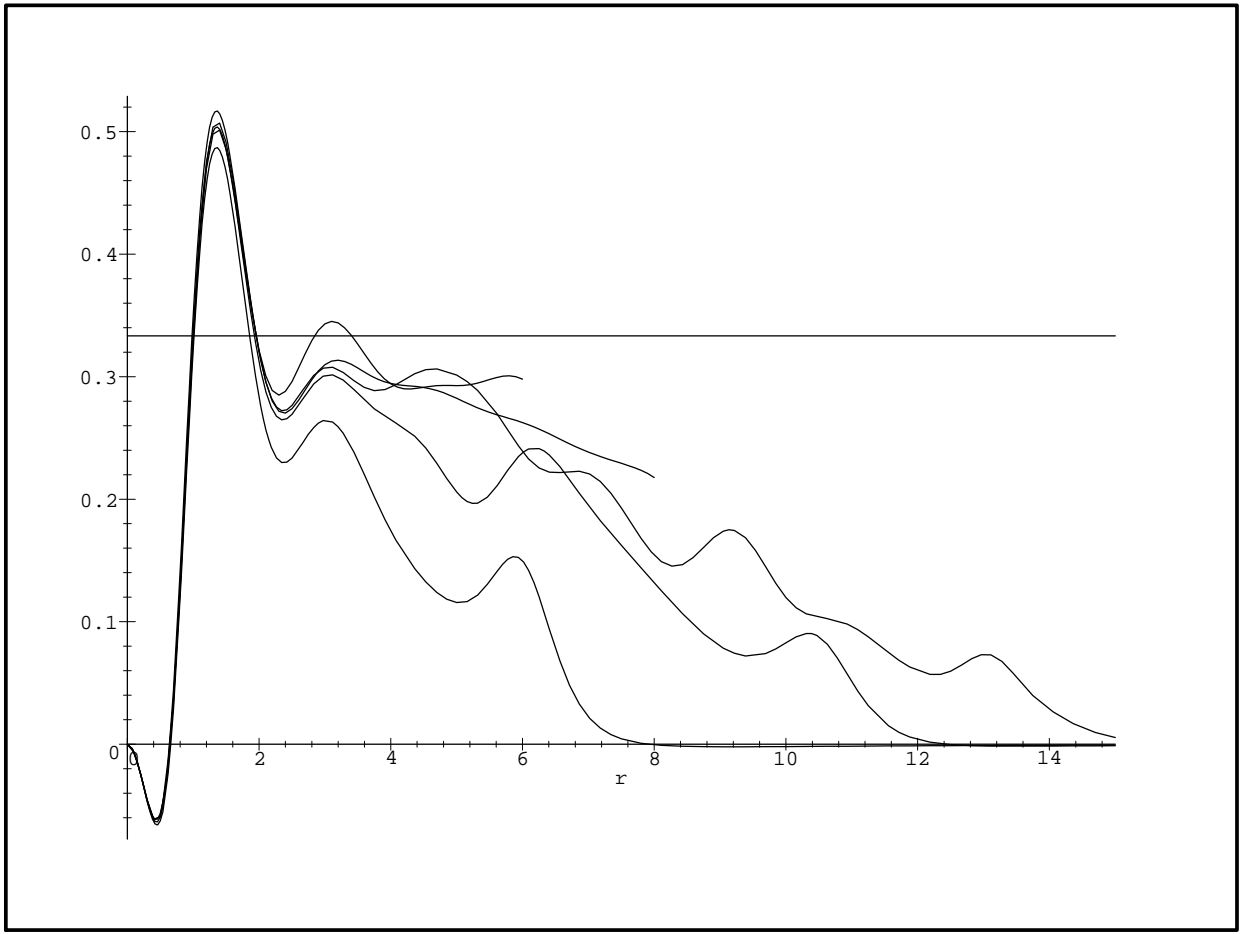,angle=270,height=14cm}}
\end{center}
\caption[]{
The quantity $\nu^{\qe}$, Eq.~(\ref{stqe}), to be interpreted as
the quasi-electron statistics parameter, versus $r$, half the
distance between the quasi-electrons. The lowest lying curve is for 20
electrons. Next are curves for 50 and 75 electrons.
The 100 electron curve is cut at $r=8$ and the 200 electron curve
at $r=6$, to avoid numerical problems.
The horizontal line is $1/3$.}
\label{fig8}
\end{figure}

These functions are plotted in Fig.~\ref{fig6}, for
$\nu=1/m=1/3$, and compared to the statistics parameter found from
the Laughlin quasi-hole states.

The exclusion statistics parameter, which is given by the asymptotic
behaviour of the two-anyon wave function as $|u|\to 0$, is
$\nu_1=\nu$ for the coherent state as well as for the projected
position eigenstate.

In the case of anyons of negative charge $-e/m$,
the relative angular momentum eigenvalues are
$-2k+\nu$
with $k$ integer, and the relative angular momentum
eigenstates have the asymptotic form
\beq
\langle u\mid k,\nu\rangle\propto (u^*)^{2k-\nu}
\eeq
as $|u|\to 0$. Thus, the state $|k,\nu\rangle$ is
non-singular if $2k-\nu\geq 0$ and normalizable if $2k-\nu>-1$.

\begin{figure}[htb]
\begin{center}
\hspace*{-9mm}
\mbox{\psfig{figure=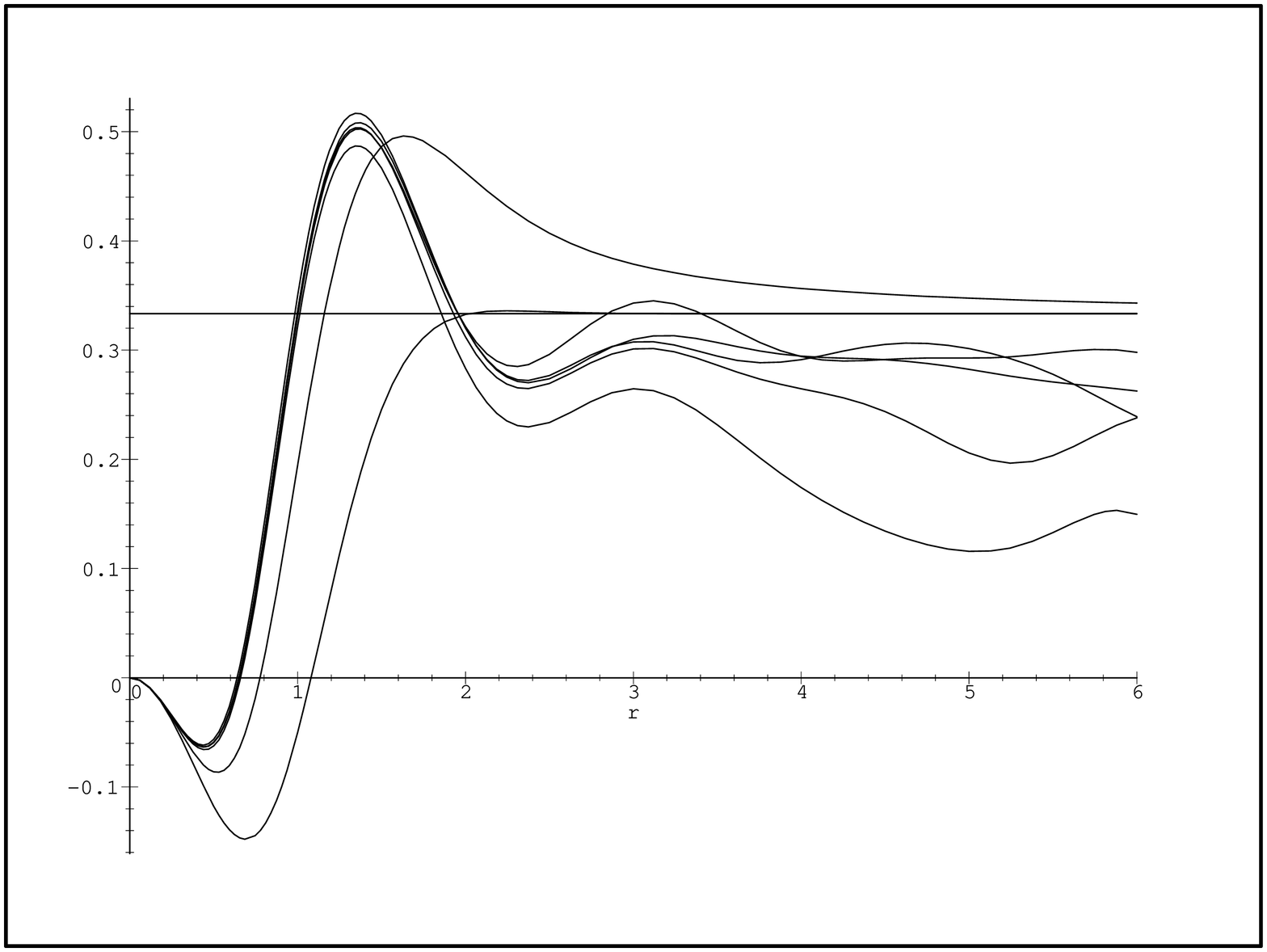,angle=270,height=14cm}}
\end{center}
\caption[]{
Small relative distance behaviour of $\nu^{\qe}$ for the cases 20, 50,
75, 100 and 200 electrons. Also shown are two curves corresponding to
ideal anyons. The five curves for the different electron numbers coincide
for small $r$. The 200 electron curve overshoots
the horizontal line $1/3$ at $r\approx 3$. The curve lying lowest
for small $r$ represents an anyon eigenstate projected onto the
lowest Landau level, whereas the curve going highest for large $r$
represents the coherent state of the su(1,1) algebra.}
\label{fig9}
\end{figure}

The formulae (\ref{koa}) and (\ref{anya})
still
hold with the substitutions $\nu\to-\nu$ and $z^*\to z$, and
with the restriction on the statistics parameter $\nu$
that $\nu\leq 0$, if only non-singular states are accepted,
or that $\nu<1$, if singular but normalizable states are accepted.
A further consequence of the substitution $z^*\to z$ is a change
of signs in the Berry phase formulae (\ref{kober}) and (\ref{anber}).
The functions $-C_{-\nu}(r)$ and $-A_{-\nu}(r)$,
with $\nu=1/3$, are plotted in Fig.~\ref{fig9} for comparison to
the statistics phase as computed for the Laughlin quasi-electrons.

The exclusion statistics parameter, given by the asymptotic
behaviour in the limit $|u|\to 0$, is now seen to be
$\nu_1=-\nu$.

\newpage

\begin{thebibliography}{99}

\bibitem{laughlin} R.B.\ Laughlin,
{\em Phys.\ Rev.\ Lett.} {\bf 50} (1983) 1395.

\bibitem{haldane} F.D.M.\ Haldane,
{\em Phys.\ Rev.\ Lett.} {\bf 51} (1983) 605.

\bibitem{halperin} B.I.\ Halperin,
{\em Phys.\ Rev.\ Lett.} {\bf 52} (1984) 1583.

\bibitem{jonm_jan} J.M.\ Leinaas and J.\ Myrheim,
{\em Nuovo Cimento} {\bf 37} (1977) 1.

\bibitem{arovas} D.\ Arovas, R.\ Schrieffer and F.\ Wilczek,
{\em Phys.\ Rev.\ Lett.} {\bf 53} (1984) 722.

\bibitem{berry} M.B.\ Berry,
{\em Proc.\ R.\ Soc.\ Lond.\ A.} {\bf 392} (1984) 45.

\bibitem{AharonBohm} Y.\ Aharonov and D.\ Bohm,
{\em Phys.\ Rev.} {\bf 115} (1959) 485.

\bibitem{arovas2} D.P.\ Arovas, in {\em Geometric Phases in Physics},
eds.\ A.\ Shapere and F.\ Wilczek.\\
World Scientific (1989).

\bibitem{john_can} M.D.\ Johnson and G.S.\ Canright, {\em Phys.\ Rev.}
{\bf B49} (1994) 2947.

\bibitem{haldane2} F.D.M.\ Haldane, {\em Phys.\ Rev.\ Lett.} {\bf 67}
(1991) 937.

\bibitem{laughplasma}
R.B.\ Laughlin, in {\em The Quantum Hall Effect},
eds.\ S.M.\ Girvin and R.E.\ Prange. Springer-Verlag (1987).

\bibitem{laughfrac} R.B.\ Laughlin, in {\em Fractional Statistics and
Anyon Superconductivity}, ed.\ F.\ Wilczek. World Scientific (1990).

\bibitem{jonm_hei} H.\ Kj{\o}nsberg and J.M.\ Leinaas, {\em
Int.\ J.\ Mod.\ Phys.} {\bf A12} (1997) 1975.

\bibitem{experiment} L.\ Saminadayer, D.C.\ Glattli, Y.\ Jin and
B.\ Etienne, {\em Phys.\ Rev.\ Lett.} {\bf 79} (1997) 2526.
\newline
R.\ de-Picciotto, M.\ Reznikov, M.\ Heiblum,
V.\ Umansky, G.\ Bunin and D.\ Mahalu, {\em Nature} {\bf 389} (1997) 162.

\bibitem{kivelson} S.\ Kivelson and M.\ Rocek, {\em Phys.\ Lett.} {\bf
156B} (1985) 85.

\bibitem{lang} S.\ Lang, {\em Algebra}. Addison-Wesley (1977).

\bibitem{girvin} S.M.\ Girvin and T.\ Jach,
{\em Phys.\ Rev.} {\bf B29} (1984) 5617.

\bibitem{jonm_hans_jan} T.H.\ Hansson, J.M.\ Leinaas and J.\ Myrheim,
{\em Nucl.\ Phys.} {\bf B384} (1992) 559.


\end{thebibliography}
\end{document}